\documentclass[a4paper,UKenglish,cleveref,autoref,thm-restate]{lipics-v2021}

\hideLIPIcs  %

\title{Can LLMs Produce Better Object-Oriented Designs than Human-Involved Development?} %

\titlerunning{Can LLMs Produce Better OOD?}

\author{Zushuai Zhang}{University of Auckland, Auckland, New Zealand}{derek.zhang@auckland.ac.nz}{}{}

\author{Elliott Wen}{University of Auckland, Auckland, New Zealand}{elliott.wen@auckland.ac.nz}{}{}

\author{Ewan Tempero}{University of Auckland, Auckland, New Zealand}{e.tempero@auckland.ac.nz}{}{}

\authorrunning{Z. Zhang, E. Wen, and E. Tempero}

\Copyright{Zushuai Zhang, Elliott Wen, and Ewan Tempero}



\nolinenumbers %

\usepackage{booktabs,tabularx,array,url}
\usepackage{graphicx}
\usepackage{xcolor}

\usepackage{tikz}

\newcommand{\stepmark}[1]{%
  \tikz[baseline=(char.base)]{
    \node[
      circle,
      draw=black,
      fill=black,
      text=white,
      inner sep=0.3pt,
      font=\sffamily\bfseries\footnotesize
    ] (char) {#1};
  }%
}

\begin{document}

\maketitle

\begin{abstract}

\textbf{Background:} Large Language Models (LLMs) are increasingly used for code generation. However, their ability to generate multi-class projects that require object-oriented design (OOD) remains unclear, especially relative to projects developed with human involvement. \textbf{Aims:} The primary objective of this study is to compare OOD quality in projects from three authorship conditions: \textit{PreAI} (human-involved projects produced before widespread LLM use), \textit{PostAI} (human-involved projects produced after widespread LLM use), and \textit{PureAI} (projects generated end-to-end by contemporary LLMs). \textbf{Method:} We conducted a comparative case study on a postgraduate Java assignment. Two offerings of the same assignment were selected as the PreAI and PostAI datasets. PureAI projects were generated using three contemporary LLMs. We analyzed OOD quality using project-level OOD metrics, code smell density, and domain modeling. \textbf{Results:} Relative to human-involved projects, PureAI projects show lower code smell density and generally appear simpler in terms of total size, complexity, and coupling. However, this is consistent with oversimplification, as it is associated with missing abstractions and weaker responsibility separation. PostAI is closer to PureAI than PreAI on many OOD measures and also shows tendencies toward oversimplification. \textbf{Conclusions:} Our findings indicate that appropriate human guidance on object-oriented decomposition and responsibility assignment remains important when LLMs are used for object-oriented design.

\end{abstract}

\section{Introduction}\label{sec:intro}

Large Language Models (LLMs) are rapidly reshaping software development \cite{hou2024large}. Although initially used mainly for small-scale assistance, such as implementing utility functions, LLMs are now increasingly capable of generating large portions of software and have also been studied for producing entire projects from natural-language descriptions \cite{du2024evaluating,liu2025projecteval}. This shift raises an important question that goes beyond functional correctness: how maintainable are the resulting projects compared with code produced through human-involved development? This question is important because software maintenance often constitutes a substantial share of lifecycle cost, and code with low maintainability can make maintenance activities more difficult and expensive \cite{chen2009empirical}. In this paper, we focus on one factor that influences maintainability: object-oriented design (OOD) quality, that is, the degree to which the organization of classes and the relationships among them align with OOD principles and heuristics \cite{chidamber1994metrics, marinescu2005measurement, booch2008object}.

Prior work has compared human-written and LLM-generated code on the same task in terms of maintainability. However, the settings used in these studies may not fully capture OOD quality in realistic software development contexts. Much of this literature focuses on problem-level programs \cite{santa2025llm,jamil2025can}, which involve solving a single programming problem, typically within a single method or class. In contrast, project-level software development requires developers to address multiple programming problems associated with different features and to organize code across interacting classes \cite{leich2005tool,apel2009overview}. OOD principles and heuristics such as low coupling and high cohesion are more meaningfully assessed at the project level, where the presence of multiple classes, methods, and fields allows structural relationships to emerge \cite{chidamber1994metrics,henderson1996coupling,marinescu2005measurement}. Additionally, most prior studies consider only human-written code produced before the widespread adoption of LLMs \cite{santa2025llm,jamil2025can}. In the LLM era, LLM-based coding assistants have become increasingly mainstream \cite{brown2025exploring}. Including a human-authored condition produced during this period is therefore analytically meaningful for assessing the potential impact of LLM use on OOD quality.

Studies have also explored LLMs in automated modeling \cite{xiao2026oodeval,de2024evaluating}. However, they primarily evaluate high-level design artifacts such as UML diagrams, which provide a limited view of whether OOD principles and heuristics are reflected in implemented code.

As a step toward addressing these gaps, we conduct a comparative case study of submissions to the same Java project from two offerings by postgraduate students. We examine three authorship conditions: PreAI, representing human-involved projects produced before widespread LLM use; PostAI, representing human-involved projects produced after widespread LLM use; and PureAI, representing projects generated end-to-end by contemporary LLMs. PureAI projects are generated using three prompt versions that provide different levels of design guidance, since prompting is one of the main mechanisms through which developers can influence LLM-generated solutions in practice \cite{tafreshipour2025prompting}, and different levels of guidance may lead to different OOD outcomes. Specifically, this study addresses the following research questions (RQs):

\begin{description}
  \item[\textbf{RQ1}] What differences exist in OOD quality among PreAI, PostAI, and PureAI projects?
  \item[\textbf{RQ2}] What effect does the level of design guidance in prompts have on the OOD quality of PureAI projects?
\end{description}

To answer RQ1, we compare PreAI, PostAI, and PureAI projects using established OOD metrics, code smell density, and a manual assessment of the extent to which implemented classes represent domain concepts expressed in the functional requirements. OOD metrics provide direct measurement of design characteristics such as coupling and cohesion. The presence of code smells and the absence of certain domain concepts both indicate potential violations of OOD principles and heuristics \cite{fowler1999refactoring, yamashita2013developers, sharma2018survey, larman2012applying}. To answer RQ2, we compare PureAI projects generated under different levels of design guidance in the prompt and examine how prompt specificity affects these same measures.

In summary, our contributions are as follows:
\begin{itemize}
    \item To the best of our knowledge, we design and conduct the first comparative analysis of OOD quality in a project-level setting across three authorship conditions: \textit{PreAI}, \textit{PostAI}, and \textit{PureAI}. To support this comparison, we provide a comprehensive, multi-dimensional evaluation that combines OOD metrics, code smell density, and requirement-based analysis of domain modeling.

    \item We provide empirical evidence that PureAI projects often appear simpler and less smell-prone than human-involved projects, but this pattern is often associated with oversimplification, reflected in fewer explicitly represented domain concepts. We also show that PostAI is closer to PureAI than PreAI on many OOD measures and therefore also shows tendencies toward oversimplification. More specific OOD-guidance prompts generally improve domain concept representation in PureAI projects without closing the gap with human-involved projects. These findings provide valuable insights into the strengths and limitations of LLMs in object-oriented settings. They also offer actionable guidance and practical implications for the effective selection and use of LLMs in object-oriented development and for more systematic code review in LLM-assisted coding environments, benefiting both practitioners and organizations. Researchers can build on our study design and findings to further study how prompting can improve design quality.
\end{itemize}

\section{Related Work} \label{sec:relatedwork}

Research on LLM-generated code initially focused on functional correctness, typically measured by executable tests. Much of this work relies on problem-level datasets such as HumanEval \cite{chen2021evaluating} and APPS \cite{hendrycks2021measuring}, where each instance is a single, isolated programming task. More recently, project-level datasets such as DevBench \cite{li2024devbench} and ProjectEval \cite{liu2025projecteval} have been introduced. These datasets contain tasks that require generating whole projects, which more closely resemble realistic software development scenarios. However, to be useful for developers, code must satisfy not only functional correctness but also non-functional requirements such as maintainability, reliability, and security \cite{siddiq2024quality}. Accordingly, recent studies have begun to evaluate these broader quality attributes in LLM-generated code \cite{liu2024no, tihanyi2025secure, siddiq2024quality}. A key limitation of this line of work is that it typically evaluates only LLM-generated code, without a human baseline.

More recent work has directly compared LLM-generated and human-written code to better understand how LLM-generated code differs from human-written alternatives. Molison et al. \cite{santa2025llm} compare human-written and GPT-4o-generated solutions on the APPS dataset and find that LLM-generated code has fewer bugs and lower fix effort, though structural issues absent from human-written code still appear in high-difficulty tasks. Jamil et al. \cite{jamil2025can} compare 984 human-written and LLM-generated (GPT-3.5-Turbo and GPT-4) code samples from the HumanEval dataset on maintainability, readability, and security. They find that LLM-generated code does not consistently outperform human-written code across all quality metrics, though GPT-4 with enhanced prompting performs best on several measures. Licorish et al. \cite{licorish2025comparing} compare human and GPT-4 Python solutions on 72 software-engineering tasks and report that human-written code adheres better to coding standards, and GPT-4 code tends to be more complex. Cotroneo et al. \cite{cotroneo2025human} examine more than 500,000 Python and Java samples and show that LLM-generated code is generally simpler, but also more likely to contain unused code or hardcoded debugging code. By contrast, human-written code generally exhibits higher complexity and more maintainability issues.

One limitation of these studies is that they typically do not include a human-authored condition produced after the widespread adoption of LLMs (i.e., PostAI). Another is that most are still based mainly on problem-level datasets such as HumanEval, leaving a gap with respect to maintainability concerns in more realistic software development settings. As a result, they rarely assess the OOD quality of code directly. Related work on LLM modeling ability \cite{xiao2026oodeval,de2024evaluating} similarly focuses on comparing LLM-generated UML diagrams with reference solutions, rather than evaluating the OOD quality of code implementations.

Another line of work compares human-written code produced with and without LLM assistance \cite{paradis2025much, xu2025ai}. These studies suggest that AI assistance can improve productivity, especially for less-experienced developers, but may also increase review and rework burdens for more experienced ones. However, this literature focuses mainly on productivity, effort, and review burden rather than on the quality of the resulting code, and it does not compare AI-assisted human-written projects with fully LLM-generated projects (i.e., PureAI).

Overall, prior studies comparing the quality of human-written and LLM-generated code mainly rely on problem-level datasets and typically omit a PostAI condition. Studies comparing AI-assisted and unassisted human-written code are closer to real development practice, but they usually emphasize downstream indicators rather than code quality and typically omit a PureAI condition. More broadly, design quality remains underexamined: studies that analyze implemented code rarely assess OOD quality directly, while studies that focus on design quality tend to evaluate higher-level artifacts such as UML diagrams rather than implemented code. Our study addresses these gaps by comparing all three conditions (PreAI, PostAI, and PureAI) in a project-level dataset and conducting a comprehensive analysis of OOD quality.

\section{Background} \label{sec:background}

\subsection{Kalah Board Game}

Kalah is a two-player board game.\footnote{https://en.wikipedia.org/wiki/Kalah} In the traditional version, each player controls six houses and one store, and each house starts with four seeds. Houses and stores are collectively called pits and are arranged on a rectangular board, with the stores at opposite ends and the houses along the sides. Players take turns sowing seeds around the board, aiming to collect more seeds than the opponent.

Implementing Kalah is the first Java assignment in a postgraduate software maintainability course. The assignment is individual and is usually due in Week 2 or Week 3. Students are required to implement the functional requirements and pass 19 test cases, each representing a complete execution of the game. They are also expected to aim for a maintainable design. As the course prerequisites include prior Java courses with project experience, students are expected to be capable of producing a complete OOD.

\subsection{Object-Oriented Design Metrics}

OOD metrics provide a quantitative basis for assessing design quality by capturing structural characteristics such as complexity, coupling, cohesion, inheritance, and size. Although size is not traditionally treated as a design characteristic in the same sense as the others, it is still important because it reflects the scale of the project. In this study, the following six metrics are used to capture these characteristics \cite{chidamber1994metrics, henderson1996coupling}:

\begin{itemize}

    \item \textbf{Weighted Methods per Class (WMC)} is the sum of McCabe's cyclomatic complexity \cite{mccabe1976complexity} over all methods in a class, where a method's cyclomatic complexity is the number of linearly independent paths through it. Higher values indicate greater class complexity.
    
    \item \textbf{Coupling Between Object Classes (CBO)} measures the number of distinct classes to which a given class is coupled. Higher values indicate stronger coupling.
    
    \item \textbf{Lack of Cohesion in Methods (LCOM)} \cite{henderson1996coupling} measures cohesion based on method--field access. For each field, it computes the proportion of methods that do not access that field, and then averages this proportion across all fields in the class. The metric ranges from 0 to 1, where higher values indicate lower cohesion.
    
    \item \textbf{Depth of Inheritance Tree (DIT)} measures the inheritance depth of a class in the inheritance hierarchy. Higher values indicate deeper inheritance.

    \item \textbf{Number of Classes ($\boldsymbol{\#}\mathbf{Cl}$)} is the total number of classes in the project, including interfaces and abstract classes.
    
    \item \textbf{Lines of Code (LOC)} is the total number of source code lines in a class, excluding comments and blank lines.

\end{itemize}

\subsection{Code Smells}

Code smells are symptoms in source code that may indicate violations of OOD principles and heuristics \cite{fowler1999refactoring, yamashita2013developers}. They can be classified into two levels of granularity: method-level and class-level smells, depending on whether they mainly affect the design of methods or classes \cite{lanza2007object, martins2021code}. Accordingly, both levels may indicate potential design issues, but class-level smells relate more directly to class design, whereas method-level smells can be more closely associated with localized implementation issues.

\subsection{Domain Concept}

A domain concept is a key entity in the application domain, and nouns in the requirements provide candidates for identifying it \cite{arora2016extracting, nanduri1995requirements}. In OOD, it is important that domain concepts be represented appropriately in the design, since missing or inadequately represented concepts may reflect inappropriate abstraction, thereby reducing comprehensibility and increasing future maintenance effort \cite{sharma2018survey,larman2012applying}.

\section{Study Design}

\subsection{Operationalization of Object-Oriented Design Metrics}\label{sec:ood-metrics}

To characterize the overall structure of each design, we measure $\#\mathrm{Cl}$, WMC, LOC, LCOM, CBO, and DIT. $\#\mathrm{Cl}$ is measured at the project level. The remaining metrics are first measured for each class and then aggregated to the project level.

For WMC, LOC, and CBO, we define three design-level measures for each metric: the sum ($\Sigma$), the average ($\mu$), and the max-percentage ($\%$). The sum captures the overall amount of the corresponding property in a design, whereas the average captures its typical class-level value. Both are computed over all classes in a project. The max-percentage is the maximum class-level value divided by the sum of that metric over all classes, expressed as a percentage. This measure reflects the proportion of the total metric value contributed by the class with the highest value, and therefore indicates whether a single class accounts for a disproportionately large share of the design with respect to complexity, size, or coupling. These yield $\mathrm{WMC}_{\Sigma}$, $\mathrm{WMC}_{\mu}$, $\mathrm{WMC}_{\%}$, $\mathrm{LOC}_{\Sigma}$, $\mathrm{LOC}_{\mu}$, $\mathrm{LOC}_{\%}$, $\mathrm{CBO}_{\Sigma}$, $\mathrm{CBO}_{\mu}$, and $\mathrm{CBO}_{\%}$.

For LCOM, we use only the average and maximum values, denoted as $\mathrm{LCOM}_{\mu}$ and $\mathrm{LCOM}_{\max}$. Because LCOM is a normalized measure of class-level cohesion, summing it across classes does not yield an interpretable quantity. The average reflects the typical class-level cohesion in the design, whereas the maximum reflects worst-case cohesion, i.e., the cohesion of the least cohesive class. For DIT, we use only the maximum value, denoted as $\mathrm{DIT}_{\max}$, which captures the deepest inheritance level in the design. Summing it across classes is not interpretable as a design-level measure, since classes in the same hierarchy would lead to overlapping counts. We also do not use the average because many classes may not participate in inheritance, making it less representative of the inheritance structure.

In total, this yields 13 metrics: $\boldsymbol{\#}\mathbf{Cl}$, $\mathbf{WMC}_{\boldsymbol{\Sigma}}$, $\mathbf{WMC}_{\boldsymbol{\mu}}$, $\mathbf{WMC}_{\boldsymbol{\%}}$, $\mathbf{LOC}_{\boldsymbol{\Sigma}}$, $\mathbf{LOC}_{\boldsymbol{\mu}}$, $\mathbf{LOC}_{\boldsymbol{\%}}$, $\mathbf{CBO}_{\boldsymbol{\Sigma}}$, $\mathbf{CBO}_{\boldsymbol{\mu}}$, $\mathbf{CBO}_{\boldsymbol{\%}}$, $\mathbf{LCOM}_{\boldsymbol{\mu}}$, $\mathbf{LCOM}_{\boldsymbol{\max}}$, and $\mathbf{DIT}_{\boldsymbol{\max}}$.

\subsection{Operationalization of Code Smells}\label{sec:smell-metrics}

In this study, we consider 11 method-level and 19 class-level smell types. They were selected to cover both granularities and to include smells that have received considerable empirical attention in the software maintenance literature, are commonly observed in codebases, and are especially harmful to maintainability \cite{kaur2020systematic, zakeri2023systematic, palomba2018diffuseness}, while also reflecting the detection capabilities of the tools described in Section~\ref{sec:metric-extraction}.

We operationalize code smells as project-level smell-instance density measures, normalizing smell-instance counts by project size to account for size differences. Project size is measured in thousands of lines of code (KLOC), computed as $\frac{\mathrm{LOC}_{\Sigma}}{1{,}000}$.

We define three code smell metrics: one overall metric, \textit{Code Smell Density} ($\mathbf{CS}_{\boldsymbol{d}}$), computed as the total number of detected smell instances divided by KLOC; \textit{Method-Level Code Smell Density} ($\mathbf{MCS}_{\boldsymbol{d}}$), computed as the number of detected method-level smell instances divided by KLOC; and \textit{Class-Level Code Smell Density} ($\mathbf{CCS}_{\boldsymbol{d}}$), computed as the number of detected class-level smell instances divided by KLOC.

\subsection{Operationalization of Concept Representation and Runtime Conformance}\label{sec:concept-metric}

This study focuses on requirement-derived concepts that are suitable for explicit modeling as Java classes. We do not analyze finer-grained elements such as fields or methods, because the requirements do not prescribe them strictly, and multiple reasonable designs are possible. We distinguish between \textit{concept representation} and \textit{runtime conformance}. Concept representation concerns whether a concept is explicitly represented as a class. Runtime conformance concerns whether the number of objects instantiated from each representing class in one execution matches the quantity implied by the requirements. Deviations from the required quantity indicate that, although the concept is modeled as a class, it may not faithfully represent the intended domain behavior and may further reduce comprehensibility. These two measurements provide a consistent basis for comparison across different designs.

From the Kalah functional requirements, we extracted nouns and identified six domain concepts suitable for explicit modeling as classes, because each has its own state, behavior, and relationships with other concepts \cite{iso19502}: \textit{Board} (Bo), \textit{Game} (Ga), \textit{Player} (Pl), \textit{Pit} (Pi), \textit{House} (Ho), and \textit{Store} (St). For each concept, we analyzed whether it is explicitly represented as a class in the design. We also analyzed runtime conformance by checking whether the number of objects instantiated from each representing class matches the requirement-implied quantity. For one execution of the game, the expected object counts are 1 \textit{Board} object, 1 \textit{Game} object, 2 \textit{Player} objects, 14 \textit{Pit} objects, 12 \textit{House} objects, and 2 \textit{Store} objects. In this study, runtime conformance is assessed by executing the full 19-test suite. Because each test executes one complete game, a concept expected to instantiate $x$ objects per execution is expected to instantiate $19x$ objects in total.

We define 13 metrics. \textit{Number of concepts} ($\boldsymbol{\#}\mathbf{Co}$) is the number of the six identified domain concepts represented in a design, ranging from 0 to 6. Board representation ($\mathbf{Bo}_{\boldsymbol{r}}$), Game representation ($\mathbf{Ga}_{\boldsymbol{r}}$), Player representation ($\mathbf{Pl}_{\boldsymbol{r}}$), Pit representation ($\mathbf{Pi}_{\boldsymbol{r}}$), House representation ($\mathbf{Ho}_{\boldsymbol{r}}$), and Store representation ($\mathbf{St}_{\boldsymbol{r}}$) are binary metrics indicating whether a given concept is explicitly represented as a class. Board conformance ($\mathbf{Bo}_{\boldsymbol{c}}$), Game conformance ($\mathbf{Ga}_{\boldsymbol{c}}$), Player conformance ($\mathbf{Pl}_{\boldsymbol{c}}$), Pit conformance ($\mathbf{Pi}_{\boldsymbol{c}}$), House conformance ($\mathbf{Ho}_{\boldsymbol{c}}$), and Store conformance ($\mathbf{St}_{\boldsymbol{c}}$) are binary metrics indicating whether the runtime object count for the classes representing the concept matches the requirement-implied quantity. Conformance is evaluated only when the concept is represented; otherwise, the project is excluded from the conformance analysis for that concept.

Two special cases are handled in the runtime-conformance analysis. First, \textit{Pit} may be represented either as a standalone class or as a superclass of \textit{House} and \textit{Store}. If \textit{Pit} is represented but \textit{House} and \textit{Store} are not, we expect 14 \textit{Pit} objects per execution. When all three concepts are represented, \textit{Pit} is typically an abstract superclass of \textit{House} and \textit{Store}. In that case, we expect 0 \textit{Pit} objects, 12 \textit{House} objects, and 2 \textit{Store} objects. Second, a single domain concept may be represented collaboratively by multiple classes. For example, \textit{Game} may be split across an abstract superclass and one or more concrete subclasses that control different parts of the game. In such cases, some representing classes may have 0 instantiated objects, while others have object counts that match the requirement-implied quantity. We consider the concept to conform at runtime if all representing classes either have 0 instantiated objects or satisfy the required object count, and at least one representing class satisfies the requirement.

\subsection{Experimental workflow}\label{sec:experimentation-workflow}

\begin{figure*}[t]
    \centering
    \includegraphics[width=\textwidth]{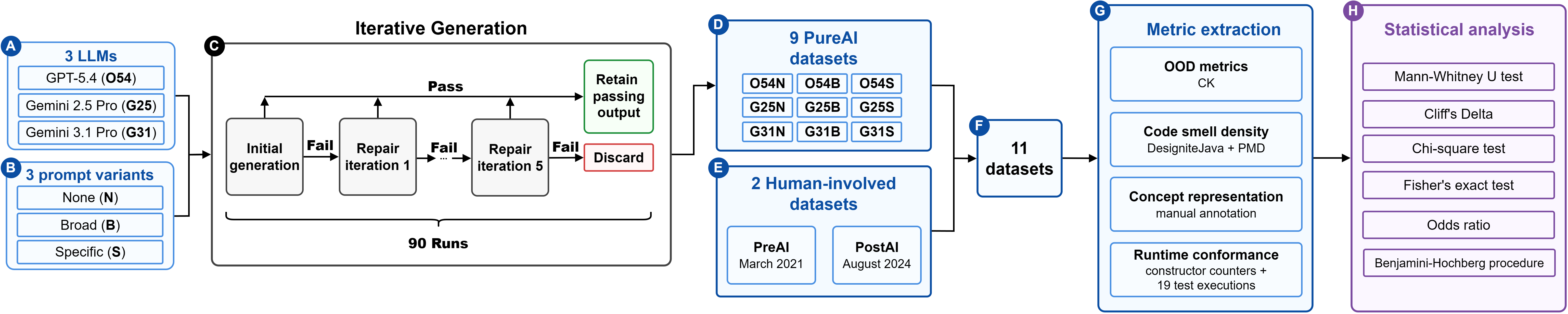}
    \caption{Overview of the experimental workflow}
    \label{fig:experimentation-workflow}
\end{figure*}

Figure~\ref{fig:experimentation-workflow} summarizes the experimental workflow, where each component is labeled with a letter. \stepmark{A} \stepmark{B} The PureAI condition was constructed by crossing three models with three prompt variants that vary in OOD-guidance specificity. \stepmark{C} \stepmark{D} For each model--prompt combination, we executed 90 independent runs, each consisting of an iterative generation process, yielding nine PureAI datasets (Section~\ref{sec:pureai-dataset}). \stepmark{E} We also included two human-involved datasets, PreAI and PostAI, corresponding to the two human-involved conditions (Section~\ref{sec:pre-post-datasets}). \stepmark{F} \stepmark{G} For all 11 datasets, we extracted the metrics defined in Sections~\ref{sec:ood-metrics}--\ref{sec:concept-metric}; the extraction procedure is described in Section~\ref{sec:metric-extraction}. \stepmark{H} Metric values were then compared across datasets using the statistical analyses described in Section~\ref{sec:data-analysis} to answer the RQs.

\subsection{PureAI dataset}\label{sec:pureai-dataset}

\begin{table}[t]
\centering
\scriptsize
\setlength{\tabcolsep}{3.5pt}
\caption{Configuration of models used in the study.}
\label{tab:model-config}
\begin{tabular}{lccccccc}
\toprule
Model & Context & Output & Cutoff & Temp. & Top-$p$ & Reasoning & Update \\
\midrule
\texttt{gpt-5.4-2026-03-05} & 1{,}050{,}000 & 128{,}000 & Aug.\ 2025 & n/a & n/a & high & n/a \\
\texttt{gemini-2.5-pro} & 1{,}048{,}576 & 65{,}536 & Jan.\ 2025 & 1.0 & 1.0 & 24{,}576 & June 2025 \\
\texttt{gemini-3.1-pro-preview} & 1{,}048{,}576 & 65{,}536 & Jan.\ 2025 & 1.0 & 1.0 & high & Feb.\ 2026 \\
\texttt{gpt-4o-2024-08-06} & 128{,}000 & 16{,}384 & Oct.\ 2023 & 1.0 & 1.0 & n/a & n/a \\
\texttt{gemini-2.0-flash} & 1{,}048{,}576 & 8{,}192 & Aug.\ 2024 & 1.0 & 1.0 & n/a & Feb.\ 2025 \\
\bottomrule
\\[-0.35em]
\multicolumn{8}{p{\linewidth}}{\footnotesize
\textit{Note.} Context = context window (tokens); Output = maximum output
tokens; Cutoff = knowledge cutoff; Temp. = temperature; Update = latest
provider-reported update; Reasoning = reasoning level or budget. For OpenAI,
we report the exact model snapshot. For Google, snapshot identifiers are not
exposed in the same way, so we report the public model alias together with the
provider-reported latest update. For \texttt{gpt-5.4},
\texttt{temperature} and \texttt{top\_p} are supported only when reasoning
effort is set to \texttt{none}; because we used \texttt{high}, these settings
were not configurable \cite{openai_using_gpt_54}. For \texttt{gemini-2.5-pro}, a thinking budget of
24{,}576 tokens corresponds to \texttt{high} reasoning effort in Google's
OpenAI-compatibility mapping \cite{google_gemini_openai_compatibility}.}
\end{tabular}
\end{table}

To construct the PureAI datasets, we selected contemporary flagship models from two providers. From OpenAI, we used \texttt{gpt-5.4-2026-03-05} (\textbf{O54}). From Google, we used \texttt{gemini-2.5-pro} (\textbf{G25}) and \texttt{gemini-3.1-pro-preview} (\textbf{G31}). We included two Google models because the latter is a preview model, whereas the former is a stable version from the same provider. The models were accessed in March 2026. Table~\ref{tab:model-config} gives the model configuration. It also includes \texttt{gpt-4o-2024-08-06} and \texttt{gemini-2.0-flash}, which were used only for the supplementary historical check described in Section~\ref{sec:pre-post-datasets}. For models that support reasoning configuration, we used the \texttt{high} setting or the closest equivalent, as this level favors complete reasoning for complex tasks where quality matters \cite{openai_reasoning_models, openai_using_gpt_54}. All generation runs used the same initial prompt:

\begin{quote}
\small\ttfamily
Implement a Java project for the Kalah game.
1. Your implementation must follow these non-functional requirements:
\texttt{\{non\_functional\_requirements\}}
2. Your imple-\\mentation must satisfy these functional requirements:
\texttt{\{functional\_requirements\}}
3. Ensure the project passes the provided test cases:
\texttt{\{test\_cases\}}
\end{quote}

\noindent
The placeholders \texttt{\{functional\_requirements\}} and \texttt{\{test\_cases\}} were instantiated with the full functional requirements and the full test suite, respectively. The placeholder \texttt{\{non\_functional\_requirements\}} had three variants: \textit{None} (\textbf{N}), \textit{Broad} (\textbf{B}), and \textit{Specific} (\textbf{S}), which differed only in OOD-guidance specificity. All three variants included the same Java-convention guidance: {\small\ttfamily ``- Follow Java coding conventions (e.g., standard Java naming conventions, meaningful identifier names, consistent formatting and indentat-\\ion, and a clean and readable structure).''} The \textit{None} variant added no further OOD guidance. The \textit{Broad} variant added only high-level OOD guidance: {\small\ttfamily ``- Use a clear object-orient-\\ed design that fits the functional requirements.''} The \textit{Specific} variant was designed to provide more explicit but still general OOD guidance, covering several core aspects of design such as object-oriented decomposition, complexity, cohesion, and coupling:

\begin{quote}
\small\ttfamily
- Use clear object-oriented decomposition, structuring the design around domain concepts from the functional requirements and aligning classes with those concepts. - Keep complexity low by avoiding complex logic, such as deeply nested control flow and overly complex expressions. - Ensure each class has a single responsibility and each method performs one coherent task. Favor short methods and small classes. - Keep related state and behavior together within the same class to promote high cohesion, and minimize unnecessary dependencies between classes to reduce coupling.
\end{quote}

For each model--prompt combination, we executed 90 independent generation runs. We chose 90 because it is close to the size of the larger human-involved dataset (93 in PreAI; see Section~\ref{sec:pre-post-datasets}), improving statistical stability and reducing variance in comparisons. Each run began with the initial prompt. If the generated project failed any test, we allowed up to five repair iterations. Runs stopped as soon as all tests passed, and we retained only the final passing output. Runs that still failed after exhausting the repair budget were discarded. This ensured that OOD quality was assessed only after functional correctness had been established, avoiding confounding the two factors. The prompt used in each repair iteration consisted of a debugging instruction: {\small\ttfamily ``Please assist in debugging a Java project. You previously generated an implementation, but the code still contains defects. Update the code so that all te-\\sts pass.''} followed by three sections: the current implementation, the test failure output, and the initial prompt for context about the functional requirements, non-functional requirements, and test cases. This procedure yielded nine PureAI datasets, denoted as \texttt{O54N}, \texttt{O54B}, \texttt{O54S}; \texttt{G25N}, \texttt{G25B}, \texttt{G25S}; and \texttt{G31N}, \texttt{G31B}, \texttt{G31S}, where each label combines the model identifier and prompt variant. The numbers of final passing outputs included in the PureAI datasets were 90, 90, 90, 87, 83, 85, 90, 90, and 90, respectively. The corresponding numbers that passed directly from the initial prompt were 64, 48, 50, 7, 7, 4, 82, 86, and 83.

\subsection{PreAI and PostAI datasets}\label{sec:pre-post-datasets}

To construct the human-involved datasets, we used student submissions from two offerings of the same Kalah assignment. \textit{PreAI} comprises the 2021 offering, whose deadline was in March 2021, whereas \textit{PostAI} comprises the 2024 offering, whose deadline was in August 2024. We selected these two offerings because the assignment and lecturers were unchanged, while the deadlines lie on opposite sides of the transition to mainstream LLM-based coding assistance. March 2021 predates GitHub Copilot's technical preview and ChatGPT's public release \cite{timeline_openai_2026}. By contrast, August 2024 is well after these tools became generally available to individual developers. We therefore use \textit{PreAI} and \textit{PostAI} as labels tied to when the submissions were produced relative to the diffusion of LLM tools. They should not be interpreted as indicating whether students did or did not use such tools. Similar to PureAI, we retained only submissions that passed all tests, yielding 93 PreAI and 57 PostAI projects.

A potential concern is whether PostAI should still be treated as human-involved, given that models available in August 2024 might have been able to generate the project end-to-end. To assess this possibility, we ran the same generate-and-repair protocol as in the PureAI condition using two models representative of that period \cite{google_model_versions_and_lifecycle}: \texttt{gpt-4o-2024-08-06} and \texttt{gemini-2.0-flash}. The latter was used as the recommended upgrade for \texttt{gemini-1.5-pro}, which was no longer available during our experiment. For this check, we used only the least prescriptive prompt variant (\textit{None}). We executed 90 runs per model, and neither model produced a single project that passed all tests. This suggests that the PostAI dataset remains largely human-involved.

\subsection{Metric extraction}\label{sec:metric-extraction}

For each of the 11 datasets, we extracted the metrics defined in Sections~\ref{sec:ood-metrics}--\ref{sec:concept-metric}. OOD metrics were computed using the CK tool \cite{aniche_ck}. Code smells were detected using DesigniteJava \cite{designite_java} and PMD \cite{pmd_tool}.

Concept representation was assessed manually using a predefined coding guide, based mainly on class names. Exact matches, minor naming variants, and close synonyms were accepted. Runtime conformance was assessed semi-automatically. For each class representing a domain concept, we manually inserted counting code into its constructor and executed the 19 tests to obtain the number of objects created. We did not instrument enums, interfaces, or abstract classes. For enums, the number of objects created per execution is fixed and determined by the number of enum constants. For interfaces and abstract classes, the count is 0 because they cannot be instantiated directly.

\subsection{Data analysis}\label{sec:data-analysis}

For \textbf{RQ1}, we compared the two human-involved datasets (\textit{PreAI} and \textit{PostAI}) with the three PureAI datasets generated using the \textit{Specific} prompt variant, namely \texttt{G25S}, \texttt{G31S}, and \texttt{O54S}. This yielded six comparisons: \texttt{G25S-PreAI}, \texttt{G31S-PreAI}, \texttt{O54S-PreAI}, \texttt{G25S-PostAI}, \texttt{G31S-PostAI}, and \texttt{O54S-PostAI}. In addition, we compared \textit{PreAI} with \textit{PostAI}, denoted as \texttt{PreAI-PostAI}, yielding seven RQ1 comparisons in total. We used only the \textit{Specific} PureAI datasets for RQ1 because the course emphasizes maintainability, and this variant provides the strongest and fairest OOD guidance for comparison with student submissions.

For \textbf{RQ2}, we examined the effect of prompt specificity within each model. For each of the three models, we compared adjacent prompt variants: Specific vs.\ Broad and Broad vs.\ None. This yielded six comparisons in total: \texttt{G25S-G25B}, \texttt{G25B-G25N}; \texttt{G31S-G31B}, \texttt{G31B-G31N}; and \texttt{O54S-O54B}, \texttt{O54B-O54N}.

For each comparison between two datasets, denoted as \texttt{$d_1$-$d_2$}, we compared each metric described in Sections~\ref{sec:ood-metrics}--\ref{sec:concept-metric} separately.
For non-binary metrics, we used the Mann--Whitney $U$ test to assess statistical significance and Cliff's delta ($\delta$) as the effect size measure. Cliff's delta ranges from $-1$ to $1$, with $0$ indicating no dominance between the two datasets. A positive $\delta$ indicates that values from the first dataset $d_1$ in the comparison tend to be larger than those from the second dataset $d_2$, whereas a negative $\delta$ indicates the opposite. We interpret Cliff's delta using the following thresholds \cite{vargha2000critique}: negligible if $|\delta| < 0.147$, small if $0.147 \leq |\delta| < 0.33$, medium if $0.33 \leq |\delta| < 0.474$, and large if $|\delta| \geq 0.474$.

For binary metrics, we used Pearson's chi-square test when all expected counts were at least 5; otherwise, we used Fisher's exact test. As the effect size measure, we used the odds ratio (OR). Let $p$ denote the probability that the event occurs in a dataset, measured as the proportion of projects in which it occurs, where the event means that a concept is represented or conforms at runtime. The odds are defined as the probability that the event occurs divided by the probability that it does not occur, i.e., $\frac{p}{1-p}$. The odds are $\infty$ when the event always occurs ($p = 100\%$), and $0$ when the event never occurs ($p = 0\%$). OR is the odds in $d_1$ divided by the odds in $d_2$.
OR ranges from $0$ to $\infty$, with $1$ indicating equal odds in the two datasets. An OR greater than $1$ indicates higher odds in the first dataset $d_1$, and an OR less than $1$ indicates higher odds in the second dataset $d_2$. OR is $\infty$ when $d_1$ has infinite odds and $d_2$ has finite odds (OR = $\frac{\infty}{\text{finite odds}}$).
Following the thresholds suggested by Chen et al. \cite{chen2010big}, we interpret OR values as follows: negligible for $\frac{1}{1.68} < \mathrm{OR} < 1.68$, small for $\frac{1}{3.47} < \mathrm{OR} \leq \frac{1}{1.68}$ or $1.68 \leq \mathrm{OR} < 3.47$, medium for $\frac{1}{6.71} < \mathrm{OR} \leq \frac{1}{3.47}$ or $3.47 \leq \mathrm{OR} < 6.71$, and large for $\mathrm{OR} \leq \frac{1}{6.71}$ or $\mathrm{OR} \geq 6.71$.

There are three situations in which OR is not estimable and no statistical test can be performed: 1. \textbf{No Data (ND)}: one dataset has no observations. This occurs in runtime-conformance analysis when a concept is not represented by any project, so no observations are available for runtime conformance; 2. \textbf{Always Occurs (AL)}: the event always occurs in both datasets ($\frac{\infty}{\infty}$); 3. \textbf{Never Occurs (NV)}: the event never occurs in both datasets ($\frac{0}{0}$).

Multiple comparisons were corrected using the Benjamini--Hochberg (BH) procedure across all statistical tests conducted in the study. Statistical significance was determined using BH-adjusted $p$-values.

\section{Results}

The results are encoded as heatmaps of effect sizes. Color hue indicates effect direction. Red indicates an increase ($\uparrow$), meaning that the first dataset in a comparison tends to have larger values for a non-binary metric or higher odds for a binary metric ($\delta > 0$ or OR $> 1$). Blue indicates a decrease ($\downarrow$), meaning the opposite ($\delta < 0$ or OR $< 1$). Grey indicates no effect ($\delta = 0$ or OR $= 1$). Color intensity indicates effect magnitude, from negligible to large. Bold cells indicate a significant Mann--Whitney $U$, Pearson's chi-square, or Fisher's exact test, depending on metric type and expected counts, at the 0.05 level; cells marked with a dagger indicate suggestive evidence ($0.05 \leq p < 0.1$).

\subsection{RQ1: Differences among PreAI, PostAI, and PureAI}

\begin{figure*}[h]
    \centering
    \includegraphics[width=\textwidth]{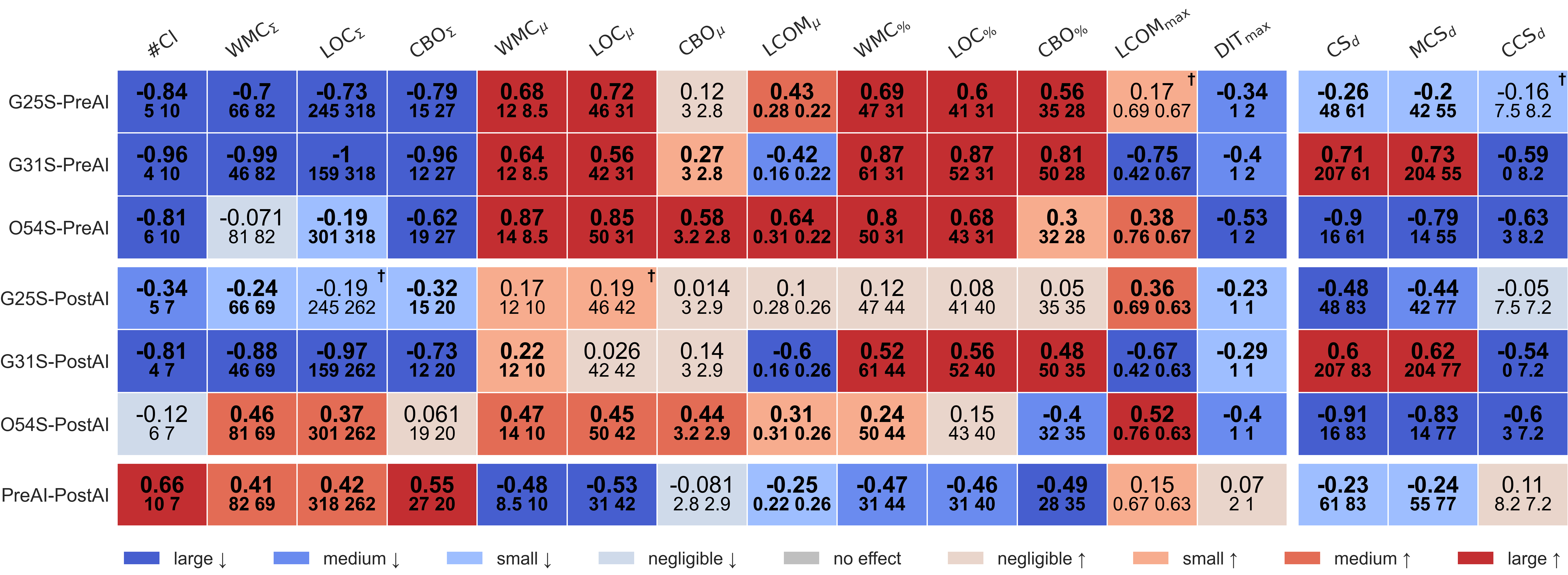}
    \caption{Cliff's delta for OOD metrics and code smell density between PreAI, PostAI, and PureAI}
    \label{fig:rq1-ood}
\end{figure*}

Figure~\ref{fig:rq1-ood} shows Cliff's delta for the OOD metrics and code smell density between PreAI, PostAI, and PureAI. Each cell reports Cliff's delta and the median metric values for the two datasets (with the ``\%'' symbol omitted for max-percentage metrics). The first six rows compare PureAI with the two human-involved datasets, with the first three against PreAI and the next three against PostAI; the last row compares PreAI with PostAI. The 16 columns comprise 13 OOD metrics followed by 3 code smell density measures. We first discuss the 13 OOD-metric columns, and then turn to the three code smell density columns.

PureAI generally has fewer classes and lower total complexity, size, and coupling, as indicated by the mostly blue cells in columns $\#\mathrm{Cl}$ to $\mathrm{CBO}_{\Sigma}$. However, in terms of per-class average and max-percentage, PureAI often shows higher values for the latter three metrics, as indicated by the mostly red cells in columns $\mathrm{WMC}_{\mu}$ to $\mathrm{CBO}_{\mu}$ and $\mathrm{WMC}_{\%}$ to $\mathrm{CBO}_{\%}$. PureAI also generally has lower average cohesion and a less cohesive worst-case class, as indicated by the mostly red cells in columns $\mathrm{LCOM}_{\mu}$ and $\mathrm{LCOM}_{\max}$ (higher LCOM indicates lower cohesion). In addition, PureAI tends to have lower maximum inheritance depth, as indicated by the blue cells in column $\mathrm{DIT}_{\max}$ (median values are at least 1 because, under the CK tool definition, all Java classes have DIT at least 1 through inheritance from \texttt{java.lang.Object}). Although these are the general tendencies, some metrics vary by model. For example, G31S shows lower $\mathrm{LCOM}_{\mu}$ ($\delta=-0.42$) and $\mathrm{LCOM}_{\max}$ ($\delta=-0.75$) than PreAI, whereas the other two models show higher values.

As shown in the last row, PreAI tends to have more classes and higher total complexity, size, and coupling, but lower average complexity and size, higher average cohesion, and lower max-percentage values for complexity, size, and coupling. Differences in $\mathrm{CBO}_{\mu}$, $\mathrm{LCOM}_{\max}$, and $\mathrm{DIT}_{\max}$ are negligible and non-significant.

An observation is that, for most OOD metrics, the PureAI-PostAI comparisons show weaker effect sizes than the corresponding PureAI-PreAI comparisons, i.e., $\delta$ is closer to 0. For example, for $\mathrm{LOC}_{\%}$, comparing rows 1 with 4, 2 with 5, and 3 with 6 shows that the effect size moves closer to 0, from 0.60, 0.87, and 0.68 in \texttt{G25S-PreAI}, \texttt{G31S-PreAI}, and \texttt{O54S-PreAI} to 0.08, 0.56, and 0.15 in the corresponding comparisons with PostAI. This indicates that PostAI is generally closer to PureAI than PreAI is. As a result, the PureAI-PostAI comparisons (rows 4--6) contain fewer significant cells than the corresponding PureAI-PreAI comparisons (rows 1--3). Also, the median values often follow either \textit{PreAI} $<$ \textit{PostAI} $<$ \textit{PureAI} or the reverse, depending on the metric. For example, for $\mathrm{LOC}_{\%}$, the median is 31\% in PreAI, 40\% in PostAI, and 41\%--52\% across the PureAI datasets.

\noindent
\textbf{\underline{Finding 1.}} Relative to human-involved projects, PureAI projects tend to have fewer classes and lower total complexity, size, and coupling, but higher average and max-percentage values of the latter three metrics, as well as lower average and worst-case cohesion. PostAI generally lies between PreAI and PureAI and is closer to PureAI than PreAI is on many metrics.

We now turn to the last three columns in Figure~\ref{fig:rq1-ood}, which report code smell density. In the first six rows, which compare PureAI with the human-involved datasets, the cells are predominantly blue, indicating that PureAI generally tends to have lower code smell density overall and at both the method and class levels. However, the pattern is not uniform across models: G31S shows increases in overall and method-level code smell density relative to both human-involved datasets, whereas the other models show decreases; G25S shows non-significant decreases in class-level code smell density relative to both human-involved datasets, whereas the other models show significant decreases. In the last row, PreAI shows significantly lower overall and method-level code smell density than PostAI. However, the difference in class-level code smell density is negligible and non-significant.

\noindent
\textbf{\underline{Finding 2.}} PureAI tends to have lower overall, method-level, and class-level code smell density than human-involved projects. PreAI also tends to have lower overall code smell density than PostAI, with the difference driven mainly by method-level code smells.

\begin{figure*}[h]
    \centering
    \includegraphics[width=\textwidth]{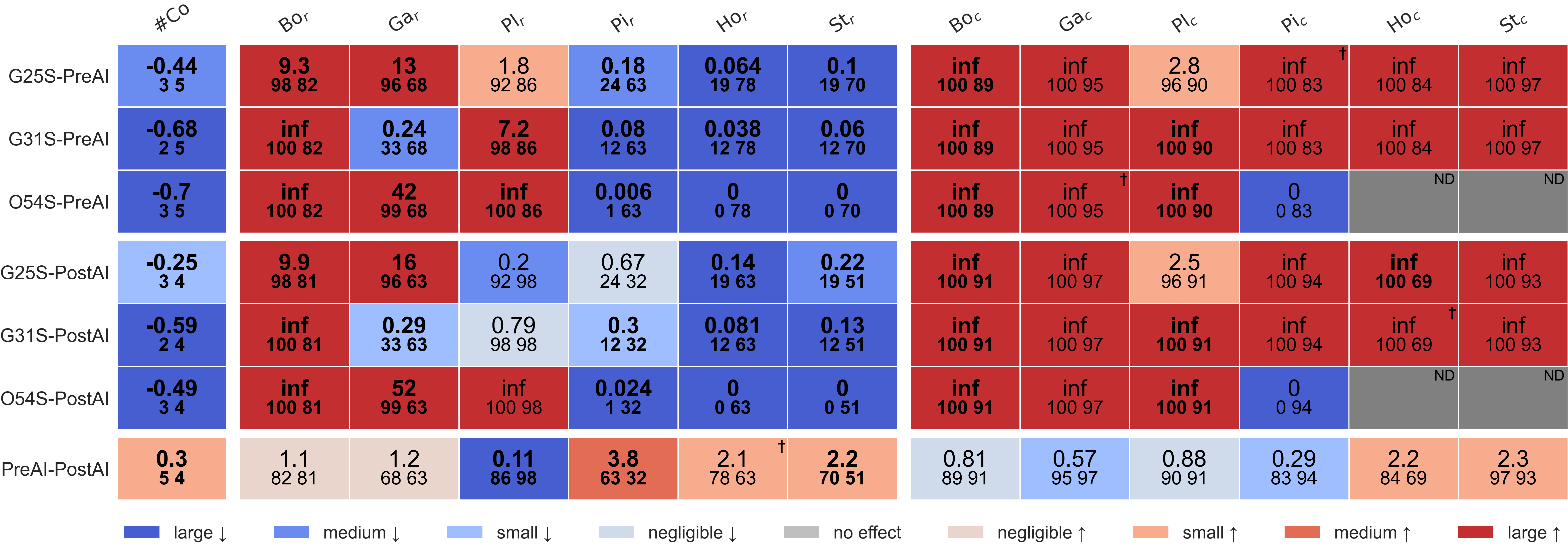}
    \caption{Cliff's delta and odds ratio for concept representation and runtime conformance between PreAI, PostAI, and PureAI}
    \label{fig:rq1-domain}
\end{figure*}

Figure~\ref{fig:rq1-domain} shows the effect sizes for concept representation and runtime conformance. The first column reports Cliff's delta for the number of represented concepts ($\#\mathrm{Co}$), together with the median values for the two datasets. The remaining 12 columns report odds ratios for concept representation and runtime conformance for each concept, together with the probability that the event occurs in both datasets (with the ``\%'' symbol omitted for probability values). Four cells are masked in dark gray, with labels at the top right referring to one of the three cases in which the odds ratio is not estimable, as defined in Section~\ref{sec:data-analysis}.

In column $\#\mathrm{Co}$, the first six rows indicate that PureAI tends to represent significantly fewer concepts than the human-involved datasets. This decrease is driven mainly by lower odds of representing \textit{Pit}, \textit{House}, and \textit{Store}, although the odds of representing \textit{Board}, \textit{Game}, and \textit{Player} are often higher. By contrast, once a concept is represented, PureAI generally has higher odds of runtime conformance, as indicated by the mostly red cells in columns $\mathrm{Bo}_{c}$ to $\mathrm{St}_{c}$. One exception is that the odds of \textit{Pit} conforming at runtime ($\mathrm{Pi}_{c}$) in O54S are lower than in the human-involved datasets, but the probability \textit{Pit} is represented ($\mathrm{Pi}_{r}$) is only 1\%.

The last row indicates that PreAI represents significantly more concepts than PostAI, driven mainly by higher odds of representing \textit{Pit}, \textit{House}, and \textit{Store}, although \textit{Player} shows lower odds. No significant difference is observed in runtime conformance.

\noindent
\textbf{\underline{Finding 3.}} Overall, PureAI projects tend to represent fewer domain concepts than human-involved projects. However, once concepts are represented, PureAI projects are generally more likely to conform at runtime. PreAI also tends to represent more concepts than PostAI, while runtime conformance does not differ significantly between them.

\subsection{RQ2: Differences among PureAI prompt variants}

\begin{figure*}[h]
    \centering
    \includegraphics[width=\textwidth]{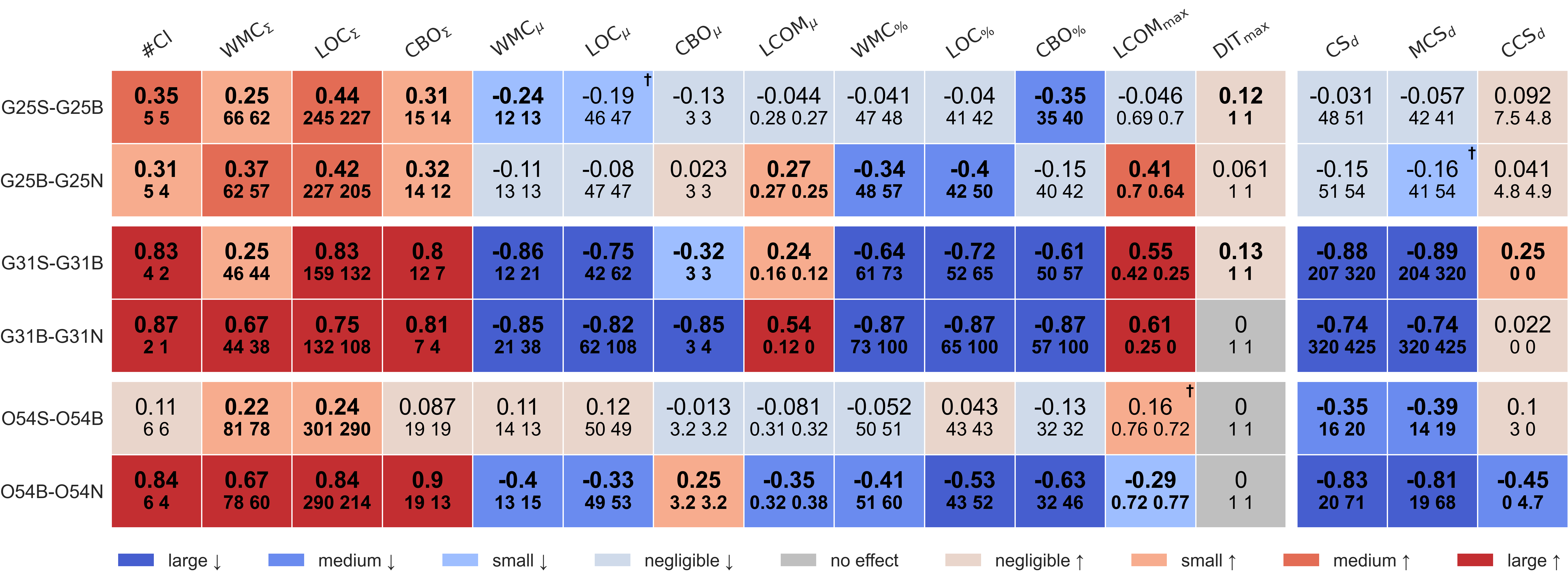}
    \caption{Cliff's delta for OOD metrics and code smell density between PureAI prompt variants}
    \label{fig:rq2-ood}
\end{figure*}

Figure~\ref{fig:rq2-ood} shows Cliff's delta for OOD metrics and code smell density between PureAI prompt variants. The six rows are grouped by model, with two prompt-variant comparisons per model. As before, we first discuss the OOD metrics and then turn to code smell density.

As OOD-guidance specificity increases from None to Broad to Specific, projects generally tend to have more classes and higher total complexity, size, and coupling, as shown by the mostly red cells in columns $\#\mathrm{Cl}$ to $\mathrm{CBO}_{\Sigma}$. In contrast, the average and max-percentage values of the latter three metrics generally tend to decrease, as shown by the mostly blue cells in columns $\mathrm{WMC}_{\mu}$ to $\mathrm{CBO}_{\mu}$ and $\mathrm{WMC}_{\%}$ to $\mathrm{CBO}_{\%}$. Worst-case cohesion ($\mathrm{LCOM}_{\max}$) also tends to decrease. Other metrics, such as $\mathrm{LCOM}_{\mu}$ and $\mathrm{DIT}_{\max}$, show less consistent overall tendencies and appear more strongly influenced by model selection. For example, $\mathrm{LCOM}_{\mu}$ increases for G31 but decreases for O54 across both prompt-specificity comparisons.

The impact of increasing OOD-guidance specificity varies across models. For example, when comparing Broad with None, 12 of 13 metrics differ significantly for both O54 and G31, but only 8 for G25. In addition, the most specific OOD guidance is not always needed for all models. When comparing Broad with None and Specific with Broad, the corresponding numbers are 12 and 2 for O54, 12 and 13 for G31, and 8 and 7 for G25. This means that both increases have comparable impacts for G31 and G25, whereas for O54, the increase from Broad to Specific has a limited impact.

\noindent
\textbf{\underline{Finding 4.}} Overall, increasing the specificity of OOD guidance tends to produce projects with more classes and higher total complexity, size, and coupling, but lower average and max-percentage values of the latter three metrics. However, worst-case cohesion tends to decrease, whereas average cohesion is more model-dependent. Additionally, moving to a more specific prompt yields limited impacts for certain models.

We now turn to the last three columns in Figure~\ref{fig:rq2-ood}, which report code smell density. As OOD-guidance specificity increases, overall and method-level code smell density generally decreases, as indicated by the mostly blue cells in $\mathrm{CS}_{d}$ and $\mathrm{MCS}_{d}$. By contrast, class-level code smell density tends to show slight increases, but these are mostly negligible and non-significant. The impact is also model-dependent. Overall and method-level code smell density show significant decreases in both comparisons for G31 and O54, whereas the decreases are non-significant and often negligible in both comparisons for G25. In addition, \texttt{O54B-O54N} is the only comparison that shows a decrease in class-level code smell density.

\noindent
\textbf{\underline{Finding 5.}} Increasing OOD-guidance specificity tends to reduce overall and method-level code smell density, while class-level code smell density generally shows negligible effects.

\begin{figure*}[h]
    \centering
    \includegraphics[width=\textwidth]{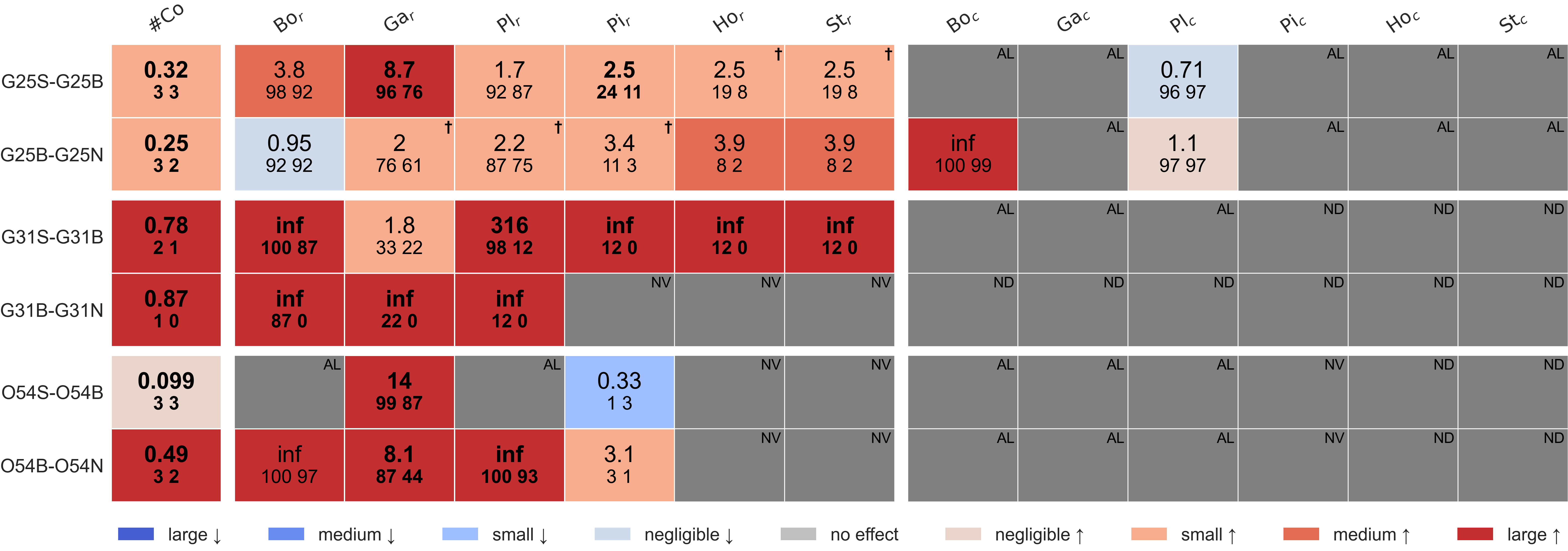}
    \caption{Cliff's delta and odds ratio for concept representation and runtime conformance between PureAI prompt variants}
    \label{fig:rq2-domain}
\end{figure*}

Figure~\ref{fig:rq2-domain} shows the effect sizes for concept representation and runtime conformance across PureAI prompt variants. As OOD-guidance specificity increases, models generally tend to represent more concepts, as shown by the mostly red cells in $\#\mathrm{Co}$. At the level of individual concepts, the odds of representing \textit{Board}, \textit{Game}, and \textit{Player} generally increase. For \textit{Pit}, \textit{House}, and \textit{Store}, the effect is more model dependent: G25 and G31 generally show higher odds of representation, whereas O54 does not.

For certain models and concepts, increasing OOD-guidance specificity does not increase the odds of representation. For G31 and O54, some comparisons for \textit{Pit}, \textit{House}, and \textit{Store} show ``NV'' (the concept is never represented in both datasets), indicating no change in odds. In other cases, a lower-specificity prompt already achieves 100\% probability of representation. For O54, \textit{Board} and \textit{Player} are always represented under both the Broad and Specific prompts, as indicated by ``AL''.

Compared with concept representation, runtime conformance appears more stable. Of the 36 runtime-conformance cells (6 comparisons $\times$ 6 columns), 13 show ``ND'', meaning that no observations are available because the concept is never represented. Among the remaining 23 cells, 18 show ``AL'', indicating that once a concept is represented, it usually conforms at runtime regardless of prompt specificity, model choice, and concept type.

\noindent
\textbf{\underline{Finding 6.}} Overall, increasing OOD-guidance specificity tends to improve concept representation, but the effect is uneven across models and concepts, and in some cases a more specific prompt yields little or no further change. By contrast, runtime conformance is generally high once a concept is represented and appears much less sensitive to OOD-guidance specificity, model choice, and concept type.

\subsection{Discussion and Implications} \label{sec:discussion}

Our results have several practical implications for the use of LLMs in project-level object-oriented development.

\noindent
\textbf{\underline{Implication 1:} PureAI projects may appear simple, but this simplicity often reflects oversimplification.} Finding 1 suggests that PureAI projects have lower total complexity, size, and coupling. However, they also tend to use fewer classes, lower inheritance depth, and fewer explicitly represented domain concepts, which suggests oversimplification. Specifically, although total complexity, size, and coupling are lower, burden is concentrated in fewer classes, average and max-percentage values increase, and cohesion often decreases because one class is more likely to represent multiple domain concepts. In practical terms, this means that fully LLM-generated projects should not be judged only by their overall simplicity. These apparent advantages may coexist with missing abstractions and weaker responsibility separation, which can hurt comprehensibility and modifiability. At the same time, Finding 2 shows that PureAI projects often have lower code smell density, and Finding 3 shows that once a concept is represented, PureAI projects are often more likely to conform at runtime. A practical implication is that human developers should retain control over high-level decomposition, especially the choice of classes and their responsibilities, while using LLMs more heavily for implementation within that structure. This helps avoid oversimplification while still leveraging the tendency of LLMs to avoid some code-level issues, such as code smells, and to implement concepts in ways that often conform at runtime.

\noindent
\textbf{\underline{Implication 2:} LLM-available development contexts may exhibit PureAI-like simplification patterns without the same cleanliness benefits.} Finding 1 shows that PostAI is closer to PureAI than PreAI on many OOD metrics. Because PureAI exhibits signs of oversimplification, this suggests projects in contexts where LLM assistance is available may also be more likely to have simpler designs that omit domain abstractions. This is further reflected in Finding 3, which shows that PostAI represents fewer domain concepts than PreAI. However, Finding 2 shows that, although PureAI tends to have lower code smell density than the human-involved datasets, PostAI has higher code smell density than PreAI, mainly due to method-level smells. Thus, PostAI shows simplification patterns similar to PureAI without showing the lower code smell density observed in PureAI. One plausible explanation is that, when LLM-generated code is not directly usable as-is, manual post-editing may make the code messier locally by introducing issues such as method-level code smells, which may be less likely to appear if the project were designed entirely by humans from the outset. This implies that productivity gains from LLM assistance~\cite{paradis2025much} may be partly offset by issues such as code smells and missing concept representation, which can increase future maintenance effort. Therefore, LLM-assisted code should be explicitly reviewed for these issues.

\noindent
\textbf{\underline{Implication 3:} Apparent improvement on one PureAI metric may be accompanied by serious trade-offs on others.} Relative to the human-involved datasets, G31S shows lower $\mathrm{LCOM}_{\mu}$ and $\mathrm{LCOM}_{\max}$, which indicate better average and worst-case cohesion, whereas the other two models show the opposite pattern (Figure~\ref{fig:rq1-ood}). However, this apparent advantage coincides with other signs of substantially weaker OOD. As shown in Figures~\ref{fig:rq1-ood} and~\ref{fig:rq1-domain}, G31S tends to have few classes (median = 4) and few represented domain concepts (median = 2). In addition, manual inspection suggests that most of its classes contain few fields, with much of the state kept in local variables. The resulting code, however, is often more step-by-step and procedural in style, which is less consistent with the object-oriented paradigm. Further, unlike the other PureAI models, G31S does not show the lower code smell density observed relative to the human-involved datasets (Figure~\ref{fig:rq1-ood}). The practical implication is that organizations should not select a model for OOD tasks based on apparent improvements in a single metric, such as cohesion. Instead, models should be evaluated systematically across multiple aspects, including OOD metrics, concept representation, and code smell density.

\noindent
\textbf{\underline{Implication 4:} More specific OOD guidance in prompts helps, but generic guidance alone may not be sufficient.} Findings 4--6 show that increasing OOD-guidance specificity generally leads to more classes, more represented concepts, lower average and max-percentage burden, and lower code smell density, especially for method-level smells. However, transitions to more specific prompts produce only limited changes for certain models. Moreover, even the most specific prompt used in this study does not fully eliminate the gap between PureAI and human-involved projects in aspects such as object-oriented decomposition. This suggests that simply making generic OOD guidance more specific may be insufficient to obtain human-comparable design quality. The practical implication is that prompt engineering alone may not be enough; as discussed in Implication 1, the more valuable input may be explicit design intent supplied by a human who understands the domain and the desired architecture (e.g., domain concepts).

\section{Threats to Validity}\label{sec:threats-validity}

The choice of measures represents a potential threat to construct validity. OOD metrics, code smell density, concept representation, and runtime conformance are informative indicators of different aspects of design quality, but they may not always align with or reliably capture developers' perceptions of quality. To reduce this threat, we report different aggregations where appropriate (e.g., sum, average, and max-percentage), and we interpret individual metrics in context rather than treating any one of them as definitive. We then analyze these complementary measures together, so that our conclusions are supported by converging evidence across multiple dimensions.

A further threat to construct validity arises from the manual assessment of concept representation. A project may represent a domain concept using a name that does not exactly match the requirements, which could lead to undercounting. We mitigate this using a predefined coding guide that accepts exact matches, minor naming variants, and close synonyms. Very poor naming may still lead to missed matches; however, this is not entirely undesirable, because one purpose of explicit concept representation is to support comprehension, and poor naming itself undermines that goal.

A threat to internal validity arises from possible differences between the maintainability guidance available to students and the design guidance given to the LLMs. We reduce this threat in two ways. First, we use the first assignment of the course, which is usually due in Week 2 or Week 3, so students have mainly been exposed to general maintainability and OOD guidance rather than advanced or course-specific design material. Second, the \textit{Specific} prompt contains guidance that is aligned with this general advice, such as clear object-oriented decomposition, single responsibility, high cohesion, reduced coupling, and low complexity. Since our evaluation also spans these same design concerns, we believe the comparison is reasonably fair. Nevertheless, we cannot rule out the possibility that students learned additional design guidance outside the course or that the LLMs benefited from hidden system instructions or prior exposure to similar human-written solutions in their training data.

Another threat to internal validity arises from uncertainty about whether differences between PreAI and PostAI are attributable to AI use. To mitigate this threat, we selected two offerings of the same assignment taught by the same lecturers, with PreAI taken from before LLM tools became generally available to the public and PostAI taken from well after that transition. Nevertheless, it remains possible that some students in the PreAI cohort had access to very early models and set them up independently, while the actual extent of LLM use in PostAI is unknown. In addition, other cohort-level factors may still differ, such as student ability, teaching-assistant support, or differences in teaching emphasis. Accordingly, the PreAI--PostAI comparison should be interpreted as observational rather than causal.

A further threat to internal validity arises from retaining only LLM outputs that passed all tests. For PureAI, runs that still failed after five repair iterations were discarded, so the analyzed outputs represent successful generations rather than the full distribution of LLM behavior. This choice avoids confounding design quality with functional correctness, but may overestimate PureAI design quality in practical use. However, the likely magnitude of this bias is limited, since only a small number of PureAI runs were discarded.

A threat to external validity arises from the generalizability of our results. The study is based on one relatively small Java assignment in a maintainability course, so the findings may generalize better to development contexts where maintainability is explicitly emphasized. They may also generalize more readily to junior or graduate-level developers, since the participants were postgraduate students with prior Java project experience. At the same time, because the task is within the capability of such developers, we do not expect the main patterns to change substantially for more experienced developers. Generalization to other programming languages, domains, or much larger projects should nevertheless be made with caution. However, even in this relatively small project, LLM outputs already show disadvantages such as missing concept representation, suggesting that these issues are unlikely to be artifacts of only small-scale systems.

\section{Conclusions and Future Work} \label{sec:conclusions}

This paper presented a comparative case study of object-oriented design quality in \textit{PreAI}, \textit{PostAI}, and \textit{PureAI} projects on the same Java assignment, evaluated through OOD metrics, code smell density, and requirement-based domain modeling. Overall, PureAI projects tend to have lower code smell density and lower total size, complexity, and coupling, but these apparent advantages are often associated with oversimplification, including fewer classes and missing abstractions. PostAI is closer to PureAI than PreAI on many OOD measures and also shows similar simplification tendencies. We further found that more specific OOD guidance generally improves abstraction in PureAI projects (more domain concepts represented), but does not fully eliminate the gap with human-involved projects. Taken together, these findings suggest that LLMs can support implementation effectively, but appropriate human guidance remains important for object-oriented decomposition and responsibility assignment.

Several directions remain for future work. First, the study can be extended to larger and more diverse project-level tasks, including different domains and programming languages, to assess the robustness of the observed patterns. Second, the requirement-based analysis can be refined beyond class-level domain modeling to examine finer-grained alignment, such as fields and methods. Third, future studies can investigate whether supplying LLMs with more explicit design intent from developers, such as expected classes and responsibilities, can reduce oversimplification and further improve OOD quality.

\bibliography{lipics-v2021-sample-article}

\end{document}